\documentstyle[12pt]{article}

\begin{document}

\begin{centering}
\bigskip
{\Large \bf Hopf-algebra description of noncommutative-spacetime symmetries}
\footnote{To appear in the
proceedings of {\it 12-th International
Colloquium ``Quantum Groups and Integrable Systems"},
Prague, 12-14 June 2003.}

\bigskip
\medskip

{\bf Alessandra AGOSTINI}\\
\bigskip
Dipartimento di Fisica, Universit\`{a} di Napoli ``Federico II'' and {\it INFN, Sez.~Napoli},\\
 Monte S.~Angelo, Via Cintia, 80126 Napoli, Italy
\end{centering}
\vspace{1cm}



\newcommand{\cplus}{\dot+}
\newcommand{\ts}{\left(}
\newcommand{\td}{\right)}
\newcommand{\qs}{\left[}
\newcommand{\gs}{\left{}
\newcommand{\gd}{\right}}
\newcommand{\qd}{\right]}
\newcommand{\ab}{{\bf a}}
\newcommand{\nn}{\nonumber}
\newcommand{\x}{{\bf x }}
\newcommand{\mA}{\mathcal{A}}
\newcommand{\mN}{\mathcal{N}}
\newcommand{\mP}{\mathcal{P}}
\newcommand{\mL}{\mathcal{L}}
\newcommand{\de}{{\mathrm{d}}}
\newcommand{\kkP}{$\kappa$-Poincar\'{e}}
\newcommand{\kM}{$\kappa$-Minkowski}
\newcommand{\Mo}{${\mathcal{M}}$}
\newcommand{\KM}{$\mathcal{M}_{\kappa}$}
\newcommand{\agg}{\triangleleft}
\newcommand{\trir}{\triangleright}
\newcommand{\R}{\mathbb{R}}
\newcommand{\C}{\mathbb{C}}
\newcommand{\cM}[1]{[M_j,{#1}_k]=i\epsilon_{jkl}{#1}_l}
\newcommand{\Da}[1]{\Delta(#1)=#1\otimes 1+1\otimes #1}
\newcommand{\Db}[1]{\Delta(#1)=#1\otimes 1+e^{-\lambda P_0^R}\otimes #1}
\def\be{\begin{equation}}
\def\ee{\end{equation}}
\def\bea{\begin{eqnarray}}
\def\eea{\end{eqnarray}}
\newcommand{\e}{\epsilon}

\newcommand{\piu}{\,\dot{+}\,}
\newcommand{\ord}[1]{:\!#1\!:}
\newcommand{\ar}{P_0,P^2}


%



%
%


\begin{abstract}
I give a brief summary of the results reported in~\cite{0306013},
in collaboration with G.~Amelino-Camelia and F.~D'Andrea.
I focus on the analysis of the symmetries of \kM\ noncommutative space-time,
described in terms of a  Weyl map.
The commutative-spacetime notion of Lie-algebra symmetries must be replaced
by the one of Hopf-algebra symmetries.
However, in the Hopf-algebra sense,  it is possible to construct an action in \kM\
which is invariant under a 10-generators Poincar\'{e}-like  symmetry algebra.
\end{abstract}

\section{Introduction}
In recent research much attention has been devoted to
the implications of noncommutativity for the classical
Poincar\'{e} symmetries of Minkowski spacetime \Mo .

For the simplest NCSTs (noncommutative spacetimes),
the canonical one ($[{\x}_\mu,{\x}_\nu]= i\theta_{\mu\nu}$),
a full understanding has been matured, and in particular it has been
established that the Lorentz-sector symmetries are broken~\cite{susskind}.
But already at the next level of complexity, the one of
Lie-algebra type ($[{\x}_\mu,{\x}_\nu]= i \zeta_{\mu,\nu}^{\sigma} {\x}_\sigma$),
our present understanding of the fate of Poincar\'{e} symmetries
is still unsatisfactory.
Some progress on this problem was reported in Ref.~\cite{0306013},
by Amelino-Camelia, D'Andrea and myself,
focusing on the illustrative example
of the ``\kM\ Lie-algebra noncommutative spacetime" \KM\ \cite{MajidRuegg,lukieAnnPhys}:
\be
\begin{array}{c}
[{\x}_j,{\x}_0]=\frac{i}{\kappa}{\x}_j~,~~~[{\x}_j,{\x}_k]=0\;\;\;j,k=1,2,3 \label{eq:kM}
\end{array}
\ee

In some mathematical studies~\cite{lukieAnnPhys,Kow02-NST}
it emerges that the symmetries of \KM\ can be described
by any one of a large number of \kkP\ Hopf algebras,
but this degeneracy (based on ``duality" axioms) remains obscure from a physics perspective.
This issue has recently taken central stage also in research
on the physical proposal~\cite{dsr1} of relativistic theories
with two invariants,
where \KM\ is being considered as a possible spacetime underlying these theories,
and the possibility of observable consequences is being
explored~\cite{grbgacgactPRDgianlufran}.

We proposed in Ref.~\cite{0306013} a new approach in which symmetries
are introduced directly at the level of the action.
We illustrated this idea in the simple case of a free scalar theory in \KM ,
and I intend to give a brief summary here of those results.

\section{Hopf-Algebra description of symmetries}\noindent
As in the familiar context of CSTs (commutative spacetimes)
one can describe an \emph{external} symmetry as a
transformation of the coordinates that
leaves invariant the action of the theory.

Let us consider the symmetry analysis for a commutative free scalar theory
\be
S(\phi)=\int{\de}^4x\;\phi(\Box - M^2)\phi\;\;\;\;\;\;(\Box=\partial_{\mu}\partial^{\mu})
\ee
The most general infinitesimal transformation generated by $T$ we can consider is:
\be
\begin{array}{ll}
x'_\mu=(1-i\epsilon T)x &\phi'(x)=\phi(x)+(x_\mu-x'_\mu)\partial^\mu\phi(x)=(1+i\epsilon T)\phi
\end{array}
\ee
Actually the action is invariant under $T$-generated transformations
if and only if
the variation of the action is zero. At the leading order in $\epsilon$:
\be
S(\phi')-S(\phi)=i\epsilon\!\int\!{\de}^4x\left(T\{\phi
(\Box-M^2)\phi\}+\phi[\Box,T]\phi\right)=0.\label{commsym}
\ee

In Minkowski spacetime
the symmetries of this action are fully described in terms of
the classical Poincar\'{e} Lie algebra ${\mathcal{P}}$,
and the operator $\Box=-P_\mu P^\mu$.

 However an algebra can be promoted to Hopf algebra introducing
some coalgebric structures\footnote{a Hopf-algebra
is an algebra endowed with a
coproduct ($\Delta: {\mathcal{A}}\to{\mathcal{A}}\otimes {\mathcal{A}}$),
a counit ($\epsilon: {\mathcal{A}}\to C$) and
an antipode ($S:{\mathcal{A}}\to {\mathcal{A}}$),
with some compatibility conditions. A trivial
Hopf-algebra is characterized by trivial structure of counit,
coproduct and antipode over the generators $T$: $\Da{T}\;\;\;\epsilon (T)=0\;\;\; S(T)=-T$}.
From this perspective ${\mathcal{P}}$
is equivalent to a ``trivial Hopf algebra".

Then for theories in CST
the symmetries can always be described in terms of a trivial
Hopf algebra. This property is connected with the
commutativity of functions. In fact, from $f {\cdot} g = g {\cdot} f$,
it follows that $\Delta$ is \emph{cocommutative} (trivial). In
general in a NCST $\Delta$ is not cocommutative, and
the Lie-algebra description cannot
be maintained, since it would not provide a sufficient set of rules
to handle consistently the laws of symmetry transformation
of products of (NC) functions.

\section{Symmetries in \kM : free scalar theory}

One can introduce the noncommutative fields $\Phi\in$\KM\ through the Weyl map.
It is well known that the Weyl map is not unique and in order to explore
the possible dependence of the symmetry analysis
on the Weyl map it is useful to
consider two explicit  choices, respectively
the "time-to-the right" $\Omega_R$ \cite{AmelinoMajid}
and the ``time-symmetrized" $\Omega_S$\cite{ALZ} map, defined in the  following way:
\bea
\Phi_{R/S}\!&\!=\!&\!\Omega_{R,S}(\phi)=\int \, \tilde{\phi}(p)
\, \Omega_{R,S}(e^{ipx}) \, {\de}^4p\\
\mbox{$\Omega_R(e^{ipx})$}\!&\!=\!&\!\mbox{$e^{i\vec{p}\vec{\x}}e^{-ip_0\x_0}\;\;\;\;\;\;\;
\Omega_S(e^{ipx}) = e^{-ip_0\frac{{\x}_0}{2}}e^{i\vec{p}\vec{\x}}e^{-ip_0\frac{{\x}_0}{2}}=\Omega_R(e^{i\vec{p}e^{-\frac{1}{2\kappa} p_0}\vec{x}-ip_0x_0})$}\nn
\eea
where $\tilde{\phi}(p)
$ 
is the inverse Fourier transform of $\phi(x)$.
Concerning the rule of integration one can adopt the "right-integral"\footnote{The
alternative definition $\int \Omega_S(\phi)=\int {\de}^4x \phi(x)$ turns out to be
equivalent~\cite{0306013}.}
$\int {\de}^4{\x}\;\Omega_R(\phi)=\int{\de}^4x\, \phi(x). $

At this point one can already formulate an educated guess for the action~\cite{0306013}
\be
S(\Phi)=\int{\de}^4{\x}\;\Phi(\Box_\kappa-M^2)\Phi\;\;\;\;\Phi\in\mbox{\KM}\label{kmaction}
\ee
where
$\Box_\kappa$ is a (differential) operator which
should reproduce
$\Box$ in the limit $\kappa\to\infty$.

By straightforward generalization of the results (\ref{commsym}),
it is natural to describe
a set of transformations $T$ as symmetries if (and only if) they close
a \emph{Hopf-algebra} structure and
\be
\int{\de}^4{\x}\;\left(T{\cdot}\left\{\Phi\left(\Box_\kappa-M^2\right)\Phi\right\}+
\Phi[\Box_\kappa,T]\Phi\right)=0 \label{twopartsnc}
\ee

The search of a maximally-symmetric
action can be structured in two steps.
In the first step one looks for a Hopf algebra whose generators $T$ satisfy
\begin{equation}
\int{\de}^4{\x}\;T\{\Phi \Box_\kappa\Phi\}=0 \label{eq:kMsym}
\end{equation}
for each differential operator $\Box_\kappa$.
In the second step one looks for an operator $\Box_\kappa$
that is invariant ($[\Box_\kappa,T]=0$) under the action of this algebra.

In introducing the concepts of translations and rotations we chose~\cite{0306013}
to follow as closely as possible the analogy with the well-established commutative
case in which:
\be
P_{\mu}(e^{ikx})=k_{\mu} e^{ikx}, \;\;\;\;M_j(e^{ikx})=-i\epsilon_{jkl}x_k\partial_l e^{ikx}
\label{eq1}
\ee
It appears natural to define translations and rotations in \KM\ by
straightforward "quantization" of their classical actions (\ref{eq1}) through the Weyl map,
but the non-uniqueness of the Weyl map does not allow to implement uniquely these definitions:
\bea
P^{R/S}_{\mu}\Omega_{R/S}(e^{ikx})
=k_{\mu} \Omega_{R/S}(e^{ikx}),&M^{R/S}_j \Omega_{R/S}(e^{ikx})
=\Omega_{R/S}(-i\epsilon_{jkl}x_k\partial_le^{ikx})&\label{eq3}
\eea
Although introduced differently (respectively in terms of
the action on right-ordered functions and on symmetrically-ordered
functions) $M_j^R$ and $M_j^S$ are actually identical. In fact
applying $M^{R/S}$ to the same element of \KM\
(for ex., $(e^{i\vec{k}\vec{\x}}e^{-ik_0{\x}_0})$)
one finds $M_j^R(e^{i\vec{k}\vec{\x}}e^{-ik_0{\x}_0})
=M_j^S(e^{i\vec{k}\vec{\x}}e^{-ik_0{\x}_0})$.
This applies also to $P_0^{R/S}$. We therefore
remove the indices $R$/$S$ for these operators. However the
ambiguity we are facing in defining spatial translations
is certainly more serious. In fact, the two candidates as
translation generators $P^{R/S}_j$
are truly inequivalent $P^R_j (e^{i\vec{k}\vec{\x}}e^{-ik_0{\x}_0})
\neq
 P^S_j (e^{i\vec{k}\vec{\x}}e^{-ik_0{\x}_0})
.
$

One can easily verify~\cite{0306013} that both
the 7-generators of operators  $(P^R_{\mu},M_j)$
and $(P^S_{\mu},M_j)$ satisfy the condition (\ref{eq:kMsym})
and do give rise to genuine Hopf algebras of translation-rotation
symmetries. In these algebras the rotations turn out to be completely
classical (undeformed) both in algebra and in co-algebra sector, whereas
for translations one finds a non-trivial coalgebra sector:
$$
\begin{array}{lll}
\Da{P_0}&\Delta P^R_j=P_j^R\otimes 1
+ e^{- \frac{P_0}{\kappa}}\otimes P^R_j &\Delta P^S_j
=P_j^S \otimes e^{\frac{P_0}{2\kappa}}+ e^{-\frac{P_0}{2\kappa}}\otimes P^S_j\nn
\end{array}
$$
Still, the action of rotations on energy-momentum
is undeformed $[M_j^{R/S},P^{R,S}_{\mu}]=i\delta_{\mu j}\varepsilon_{jkl}P^{R,S}_l$.

In including also boosts to obtain 10-generator symmetry algebras one finds that the action
on functions in \KM\ cannot be obtained by ``quantization" of the classical
action $N_j^{R/S}\Omega_{R/S}(f)=\Omega_{R/S}(i[x_0\partial_j-x_j\partial_0]f])$.
In fact $N_j^{R/S}$ do not close a consistent Hopf algebra structure:
the coproduct $\Delta(N_j^{R/S})$ is not an element of the algebraic
tensor product
of the algebra generated
by $(P^R_{\mu},M_j,N_j^R)$.
Therefore the ``classical" choice $N_j^{R/S}$ cannot be combined
with $(P^{R/S}_{\mu},M^{R/S}_j)$. However, a 10-generator
symmetry-algebra extension does exist, but it requires nonclassical boosts.

We considered~\cite{0306013} the most general form of deformed boost
generators ${\mathcal{N}}_j$ that transform as vectors under rotations
$$
\begin{array}{c}
{\mathcal{N}}_j \Omega(\phi)
=\Omega\{[ix_0A(-i\partial_x)\partial_j
+\kappa x_jB(-i\partial_x)
-\frac{x_l}{\kappa}C(-i\partial_x) \partial_{l} \partial_{j}
-i\epsilon_{jkl} x_k D(-i\partial_x)\partial_{l}]\phi\}
\end{array}
$$
where $A,B,C,D$ are unknown functions of $P^R_\mu$
(in the classical limit $A=i,D=0$; moreover, as $\kappa \rightarrow \infty$
one obtains the classical limit if $C/\kappa\rightarrow 0$
and $B \rightarrow \kappa^{-1} P_0$).

Imposing consistency of the 10-generator Hopf-algebra structure
and imposing that the classical Lorentz-subalgebra relations are preserved
one obtains some constraints
on $A,B,C,D$. The solution is
\bea
{\mN}_j^R\Omega_R(f)&=&\Omega_R([ix_0\partial_j
+\frac{\kappa}{2}x_j(1-e^{2i\frac{\partial_0}{\kappa}}-\frac{\nabla^2}{\kappa^2})
-\frac{x_l}{\kappa}\partial_l\partial_j]f)\label{boostRfin}\\
{\mN}_j^S\Omega_S(f)&=&\Omega_S([ix_0\partial_j
-x_j(\kappa\sinh(i\frac{\partial_0}{\kappa})+\frac{\nabla^2}{2\kappa})
+\frac{x_l}{2\kappa}\partial_l\partial_j]e^{i\frac{\partial_0}{2\kappa}}f)
\label{boostSfin2}
\eea
As in the case of the rotations one can easily verify that ${\mN}^R_j,{\mN}^S_j$
are equivalent, and it is therefore appropriate to remove the label $R/S$.
It is easy to verify that the Hopf algebras $(P^R_\mu,M_j,{\mN}_j)$
and $(P^S_\mu,M_j,{\mN}_j)$ both satisfy
all the requirements for a candidate symmetry-algebra
for theories in \KM . In summary we have two candidate Hopf algebras of
 10-generator Poincar\'{e}-like symmetries: $(P^R_\mu,M_j,{\mN}_j)$ is
 the well-known Majid-Ruegg bicrossproduct \kkP\ basis\cite{MajidRuegg},
 while $(P^S_\mu,M_j,{\mN}_j)$ is a new type of bicrossproduct basis
 which had not previously emerged in the literature~\cite{0306013}.


The final step is to look for a differential operator $\Box_\kappa$
suitable for a maximally-symmetric action. It is easy to verify that
the proposal $\Box_\kappa=\left(2\kappa\sinh\frac{P_0}{2\kappa}\right)^2
- e^{\frac{P_0^2}{\kappa}}P_R^2
$
satisfies $[\Box_\kappa,T]=0$ for every $T$
both in $(P^R_\mu,M_j,{\mN}_j)$) and  $(P^S_\mu,M_j,{\mN}_j)$).
Therefore, it makes the action (\ref{kmaction})
invariant both under $(P^R_\mu,M_j,{\mN}_j)$ transformations
and under $(P^S_\mu,M_j,{\mN}_j)$ transformations.

In this analysis the ambiguity associated with the choice of a Weyl
map led to consideration of two Hopf algebras,
$(P^R_\mu,M_j,N_j)$ and $(P^S_\mu,M_j,N_j)$,
which originate from two different choices
of ordering in \kM\ (in the sense codified in the Weyl maps $\Omega_R$
and $\Omega_S$).
Of course, one could consider other types of ordering
conventions. This would lead to other candidates $P^*_\mu$ as translation
generators and, correspondingly, other candidate
10-generator Hopf algebras of Poincar\'{e}-like symmetries
for \KM\ of the type $(P^*_\mu,M_j,{\mN}_j)$.

\section{More on the description of translations}

The results obtained above
were based on the  natural symmetry requirement (\ref{twopartsnc}), that however
deserves a few more comments.
Let us consider an infinitesimal translation generated by  $T=-i{\e}^{\mu}\partial_{\mu}$
(with an expansion parameter $\alpha\in R$):
\bea
{\x}\rightarrow{\x}'={\x} - \alpha \e, &&\Phi({\x})\rightarrow \Phi'({\x})
=\Phi({\x})+i\alpha T\Phi({\x})+O(\alpha^2)\nn
\eea

Following the analogy with corresponding analyses in CSTs there are actually two possible
starting points for a description of $T$ as a symmetry of the action:
\bea
I)\;\;\;\delta_{I} S(\Phi)= i \int{\de}^4{\x}\;T{\cdot}\{\Phi(\Box-M^2)\Phi\}
= 0&& II)\;\;\;\delta_{II} S(\Phi)=S(\Phi')-S(\Phi) = 0 \nn
\eea
In the context of theories in CSTs the conditions
I) II)
are easily shown to be equivalent.
But in a NCST this is not necessarily the case.
By a straightforward calculation one can see that
assuming commutative translation parameters $\epsilon$, $\delta S_I\neq\delta S_{II}$.
If one wants to preserve the double description
I) II) of symmetry under translation transformations
it is necessary~\cite{0306013}
to introduce noncommutative transformation parameters.
In fact, it is easy to verify that assuming
$
[\e_j,\x_0]=i\kappa^{-1}\e_j,\; [\e_j,\x_k]=0 \label{oeckdiff},
$
one finds that the conditions
I) II) are
equivalent.
It appears plausible that other
choices of noncommutative transformation parameters would preserve
the double description
of symmetry. But it is interesting that this choice of noncommutativity
of the transformation parameters allows to describe them as differential
forms\footnote{Note that this is one of the two differential calculi
introduced in Ref.~\cite{oeckdiff}.}, $\e_\mu=\de\x_\mu$. This connection with differential
forms leads to the following description of translations
\bea
{\x}_\mu\rightarrow{\x}_\mu'={\x}_\mu + \de\x_\mu &&\Phi({\x})\rightarrow \Phi'({\x})
=\Phi({\x})+ i\de\x_\mu P^\mu\Phi \nn
\eea
where the $\de\x_\mu$ describe the proper concept of
differential forms for \KM\ and the $P^\mu$ act as in (\ref{eq3}).
This is rather satisfactory from a conceptual perspective, since
even in CST an infinitesimal translation
is most properly described as ``addition" of a differential form.
The differentials satisfy the relations
$[\de\x_{\mu},\x_{\nu}]=i\delta_{\mu j}\delta_{\nu 0}\kappa^{-1}\de\x_j$
as required for our translations to preserve the commutators of \KM .
An infinitesimal translation $\Phi' \equiv \Phi+\de\Phi$
associates to each element of \KM\ an element of the algebra \KM $\oplus \Gamma$
defined over a vector space that is direct sum
of \KM\ and
the bimodule 
$\Gamma$, over \KM ,
with product rule
$
(\Phi+\de\Phi)(\Psi+\de\Psi)
=\Phi\Psi+\Phi{\cdot}\de\Psi+\de\Phi{\cdot}\Psi=\Phi\Psi+\de(\Phi\Psi)
$.
This algebra is isomorphic to \KM\ through the map $1+\de$.
Then an infinitesimal translation transforms an
element of \KM\ in an element of a ``second copy" of
\KM . It is a transformation internal to the
same abstract algebra.
This abstract algebra {\underline{is}} our ``space of
functions of the spacetime coordinates".

\section{Closing remarks}
We introduced~\cite{0306013} a concept of NCST symmetry,
which follows very closely the one adopted in CSTs, and
is naturally analyzed
in terms of a Weyl map.
We did find 10-generators symmetries
of a free scalar 
theory in \KM . These symmetries can be formulated
in terms of Hopf-algebra 
versions of the classical Poincar\'{e} symmetries.

The form of the commutation
relations of \KM\ clearly suggest that classical rotations can
be implemented as a symmetry, and this finds confirmation also
at the level of the analysis of the action.
Instead the $\kappa$-Minkowski commutation
relations are clearly not invariant under classical translations.
Still, we have shown that one can construct theories in \KM\ that
enjoy a deformed (Hopf-algebra) translational symmetry.
For boosts something analogous
to what happens for translations occurs: classical boosts are not
a symmetry of \KM , but, as we showed, there is
a deformed version of boosts that are symmetries.

Our analysis allowed us to clarify the nature of the ambiguity
in the description of the symmetries of theories in these NCSTs,
but it appears that we are left with a choice between different
realizations of the concept of translations.
It remains to be seen whether this ambiguity
can be removed at some deeper level of analysis.
A natural context in which to explore this issue
might be provided by attempting to construct gauge theories in \KM\ following the
approach here advocated.

\end{document}